\documentclass[%
 reprint,
superscriptaddress,
 amsmath,amssymb,
 aps, 
prd,
nofootinbib]{revtex4-2}

\usepackage{graphicx}                
\usepackage{dcolumn}
\usepackage{mwe}
\usepackage{multirow}
\usepackage{amsmath}                 
\usepackage{bm} 
\usepackage{comment}
\usepackage{float}                   
\usepackage{hyperref}                
\usepackage[margin=0.5in]{geometry}
\usepackage{aas_macros}
\usepackage{braket}
\usepackage{natbib}

\usepackage{soul}

\usepackage{xargs}
\usepackage[pdftex,dvipsnames]{xcolor}

\newcommand{\re}{\operatorname{Re}}

\newcommand{\rmax}{r_{\rm max}}

\begin{document}

\preprint{APS/123-QED}
\title{Schr\"{o}dinger-Poisson Solitons: Perturbation Theory}


\author{J. Luna Zagorac}
\email{luna.zagorac@yale.edu}
\affiliation{Department of Physics,
Yale University,
New Haven, CT 06520, USA}

\author{Isabel Sands}
\affiliation{Department of Physics,
Yale University,
New Haven, CT 06520, USA}

\author{Nikhil Padmanabhan}
\affiliation{Department of Physics,
Yale University,
New Haven, CT 06520, USA}

\author{Richard Easther}
\affiliation{Department of Physics,
University of Auckland,
Private Bag 92019,
Auckland, New Zealand}

\date{\today}

\begin{abstract}
Self-gravitating quantum matter may exist in a wide range of cosmological and astrophysical settings from the very early universe through to present-day
boson stars. Such quantum matter arises in a number of different theories, including the Peccei-Quinn axion and UltraLight (ULDM) or Fuzzy (FDM) dark matter scenarios.
We consider the dynamical evolution of perturbations to the spherically symmetric soliton, the ground state solution to the Schr\"{o}dinger-Poisson system
common to all these scenarios. We construct the eigenstates of the Schr\"{o}dinger equation, holding the gravitational potential fixed to its ground state value. We see that
the eigenstates qualitatively capture the properties seen in full ULDM simulations, including the soliton ``breathing'' mode, the random walk of the soliton center, and quadrupolar
distortions of the soliton. We then show that the time-evolution of the gravitational potential and its impact on the perturbations can be well described within the framework
of time-dependent perturbation theory. As an illustrative example, we apply our formalism to a synthetic ULDM halo. We find the soliton core accounts for approximately 30\% of the halo's wavefunction throughout its evolution, with higher modes accounting for the halo's NFW skirt, and relatively little mixing between different $\ell$ modes. Our results provide a new analytic approach to understanding the evolution of these systems as well as possibilities for faster approximate simulations.
\end{abstract}

\maketitle


\section{Introduction}

Standard Lambda Cold Dark Matter ($\Lambda$CDM) cosmology successfully describes structure formation on large scales; however, it does not necessarily account for observations on galactic and subgalactic scales. For example, CDM N-body simulations   predict dark matter halos with a central ``cusp" while many observed galaxy rotation curves are better described by ``cored" profiles with roughly constant central densities \cite{1996ApJ...462..563N, 2015PNAS..11212249W, 2019A&ARv..27....2S}.  Likewise, CDM  simulations yield more subhalos than are expected from the observed numbers of dwarf galaxies, leading to the so-called ``missing satellite" problem \cite{1993MNRAS.264..201K, 1999ApJ...522...82K, 1999ApJ...524L..19M}. Such discrepancies may be attributable to baryonic processes or even non-Newtonian dynamics \cite{Famaey:2011kh}, but may also be resolved by dark matter scenarios whose properties differ from those of simple CDM. 

One such candidate is UltraLight Dark Matter (ULDM), also known as Fuzzy Dark Matter (FDM). Consisting of an axion-like boson  with a mass between 10$^{-23}$ to 10$^{-20}$ eV,  structure formation in ULDM scenarios is suppressed on scales smaller than the corresponding de Broglie wavelength of up to a few kiloparsecs \cite{2000PhRvL..85.1158H}. ULDM  can coalesce into a Bose-Einstein condensate (BEC) whose behavior is described by a macroscopic wavefunction~\cite{2014NatPh..10..496S, 2015PhRvD..92j3513G, 2011PhRvD..84d3531C, 2011PhRvD..84d3532C} governed by the coupled Schr\"{o}dinger-Poisson  system. The ground  state solution of this system is a soliton, but the astrophysical dynamics of halo formation lead to configurations with a solitonic core embedded in a Navarro-Frenk-White (NFW)  ``skirt''  \cite{2014PhRvL.113z1302S}.

Structure formation with ULDM reproduces the successes of $\Lambda$CDM on large scales while producing cored halos and substructure that are potentially more consistent with observations on small scales~\cite{2014NatPh..10..496S,2020PASA...37....9K, 2021arXiv210111735H}. In addition to dark matter, the Schr\"{o}dinger-Poisson system of equations governing ULDM dynamics emerges in other systems of interest, including boson stars \cite{Guzman2004,Schwabe2016,mocz2017,2021PhRvD.104b3504D} and the very early universe \cite{Musoke:2019ima, 2020JCAP...07..030N, Eggemeier:2020zeg}. This motivated our study of the dynamics of the Schr\"{o}dinger-Poisson system. 

While the ground state of the Schr\"{o}dinger-Poisson system is well studied,  in most astrophysical systems, one would expect the excited states 
to be just as relevant as the ground state, given that the ``NFW skirt'' of a ULDM halo must be built up of excited states; see e.g. Refs.~\cite{2013ApJ...763...19R, PhysRevD.50.3655, PhysRevD.81.044031, PhysRevD.103.083535} . However, 
the gravitational coupling makes the system nonlinear in the wavefunction making it challenging to explore the excited 
states of this system, and most analyses have relied on directly simulating the full system.

As was pointed out in Ref.~\cite{2020arXiv201111416L}, in the limit that the density of the system is approximately constant in time, one can avoid the complications of the full system and solve the Schr\"{o}dinger equation alone, treating the fluctuations
in the density as perturbations. 
This is further helped by the fact the mapping from density to gravitational potential is a smoothing 
operation, and therefore naturally reduces the impact of small scale fluctuations. 
This paper aims to develop this idea, primarily focusing on the perturbations to the soliton as a toy example. This work is a natural continuation 
of the results presented in Ref.~\cite{2020arXiv201111416L}, although there have been a number of other explorations of perturbations in this system, eg. \cite{2019arXiv191210585G, 2021arXiv210100349S, PhysRevD.97.103523, 2019PhRvD..99f3509L}. 

Throughout this paper, we will present numerical results from a pseudo-spectral solver of the full Schro\"{o}dinger-Poisson system, {\sc chplUltra}. We developed {\sc chplUltra} based on the algorithm of {\sc PyUltraLight}: a sibling code whose specifics are discussed in detail in Ref.~\cite{2018JCAP...10..027E}. One detail in which the {\sc PyUltraLight} and {\sc chplUltra} diverge is the algorithm used for computing the potential; whereas {\sc PyUltraLight} uses Fourier transforms and periodic boundary conditions, {\sc chplUltra} utilizes a Green's function approach that allows for isolated boundary conditions. This difference, along with the implementation of {\sc chplUltra} is explained in detail in Ref.~\cite{2019SC2}. Additionally, details of {\sc chplUltra} and our code units are summarized in Appendix~\ref{sec:code-units}.

The rest of our paper is organized as follows. We review the construction of the relevant eigenstates in Section \ref{sec:method}, paying attention to the impact of 
the boundary conditions on our results. Section \ref{sec:perturbed-solitons} starts by demonstrating that perturbing a soliton by these eigenstates can 
qualitatively reproduce many of the results seen in full ULDM simulations. It then continues to show that the time evolution of these perturbations in the full 
system can be accurately captured by a simple perturbative calculation. 
In Section \ref{sec:uldm} we consider a more realistic case, and decompose a ULDM halo into its eigenstates and track their evolution. 
Finally, we discuss our results in Section \ref{sec:discussion}. 

\section{ULDM Eigenstates} \label{sec:method}

\subsection{Eigenfunction Expansion} 
\label{eq:SP}

We will be solving the  Schr\"{o}dinger-Poisson system,  
\begin{align}
  -i \hbar \frac{\partial}{\partial t} \psi &=
              \Biggl[-\frac{\hbar^2}{2m_a} \nabla^2 + m_a \Phi \Biggr] \psi \label{eq:Schrodinger}\\
      \nabla^2 \Phi &= 4 \pi G m_a \rho 
      \label{eq:poisson}
\end{align}
where $\psi$ is the ULDM wavefunction, with $\rho = |\psi|^2$ as the corresponding density
and $\Phi$ as the gravitational potential. In what follows, we work in units of $m_a = \hbar = G = c = 1$, where $m_a$ is the mass of the particle. The mapping from natural to physical units is given in Appendix~\ref{sec:code-units}.

The Schr\"{o}dinger equation is linear but the gravitational interaction introduces a nonlinear dependence on $\psi$, rendering the system substantially more challenging to solve. However, in many systems of
interest the potential is approximately constant, especially when averaged in time and over
small-scale fluctuations. This suggests  the  approximation
\begin{align}
  -i  \frac{\partial}{\partial t} \psi =
  \Biggl[-\frac{1}{2} \nabla^2 + \langle \Phi \rangle \Biggr] \psi \label{eq:Schro-avg}
\end{align}
where $\langle \Phi \rangle$ is an averaged gravitational potential that is assumed to
be constant in time. 

We expand the ULDM wavefunction at $t=0$ as
\begin{align}
\psi(t=0) = \sum_{i=1}^{N} c_i \phi_i \,
\end{align}
where the $c_i$ are 
complex expansion coefficients, $\phi_i$ are the system's eigenstates, and $N$ is a  finite truncation of the basis. If the $\phi_i$ are assumed to be orthonormal we can project out their weights
\begin{align}
c_i = \int \, d^3 r\, \psi({\mathbf r}) \phi_{i}^{*}({\mathbf r}) \,
\end{align}
where the integral is over all space. If we ignore the backreaction on the potential, the wavefunction evolves via
\begin{align}
  \psi(t) = \sum_{i=1}^{N} c_i \exp \Biggl(-i E_i t\Biggr) \phi_i \, 
\end{align}\label{eq:lin-theory}
where $E_i$ is the eigenenergy associated with state $i$.

\subsection{Construction of Eigenstates} \label{subsec:methods}

There is substantial literature on solving the Schr\"{o}dinger-Poisson (or Schr\"{o}dinger-Newton) eigensystem; see e.g.  Refs.~\cite{1995RpMP...36..331L, 1996PhLB..366...85S, 1998MPLA...13.2327B, 1999Nonli..12..201T, 2002math.ph...8045H, 2002math.ph...8046H, 2003Nonli..16..101H}.
However, since we have assumed that  $\Phi$ is constant we are effectively determining
eigenstates of the Schr\"{o}dinger equation, without the additional coupling to the Poisson equation. Furthermore, we restrict our attention to spherically symmetric potentials but allow the perturbations to break spherical symmetry.

With these assumptions we can separate variables so that the eigenstates are each products of a radial and an angular component: $\phi_{n \ell m} = f_{n \ell}(r)Y^m_{\ell}(\theta,\phi)$. Re-arranging Eq.~\ref{eq:Schro-avg} and dividing through by $Y_{\ell m}$, we arrive at
\begin{align}
  \frac{1}{r^{2}} \frac{\partial}{\partial r}\left(r^{2} \frac{\partial f_{n  \ell}}{\partial r}\right) + \frac{\ell(\ell + 1)}{r^2} f_{n \ell}  &= 2(\langle \Phi \rangle - E_{n  \ell})f_{n  \ell} \, ,
\end{align}
where $E_{n  \ell}$ is the eigenvalue of eigenstate $f$. The substitution $u_{n\ell} = r f_{n\ell}$ transforms the above equation into
\begin{align}
\frac{\partial^2 u_{n\ell}}{\partial r^2} + \frac{\ell(\ell + 1)}{r^2} u_{n\ell} (r) - 2 \langle \Phi(r) \rangle u_{n\ell} (r) &= -2 \, E_{n\ell} \, u_{n\ell}(r) \, .   \label{eq:rescale}
\end{align}

\begin{figure*}[tb]
    \centering
    \includegraphics[width=2\columnwidth]{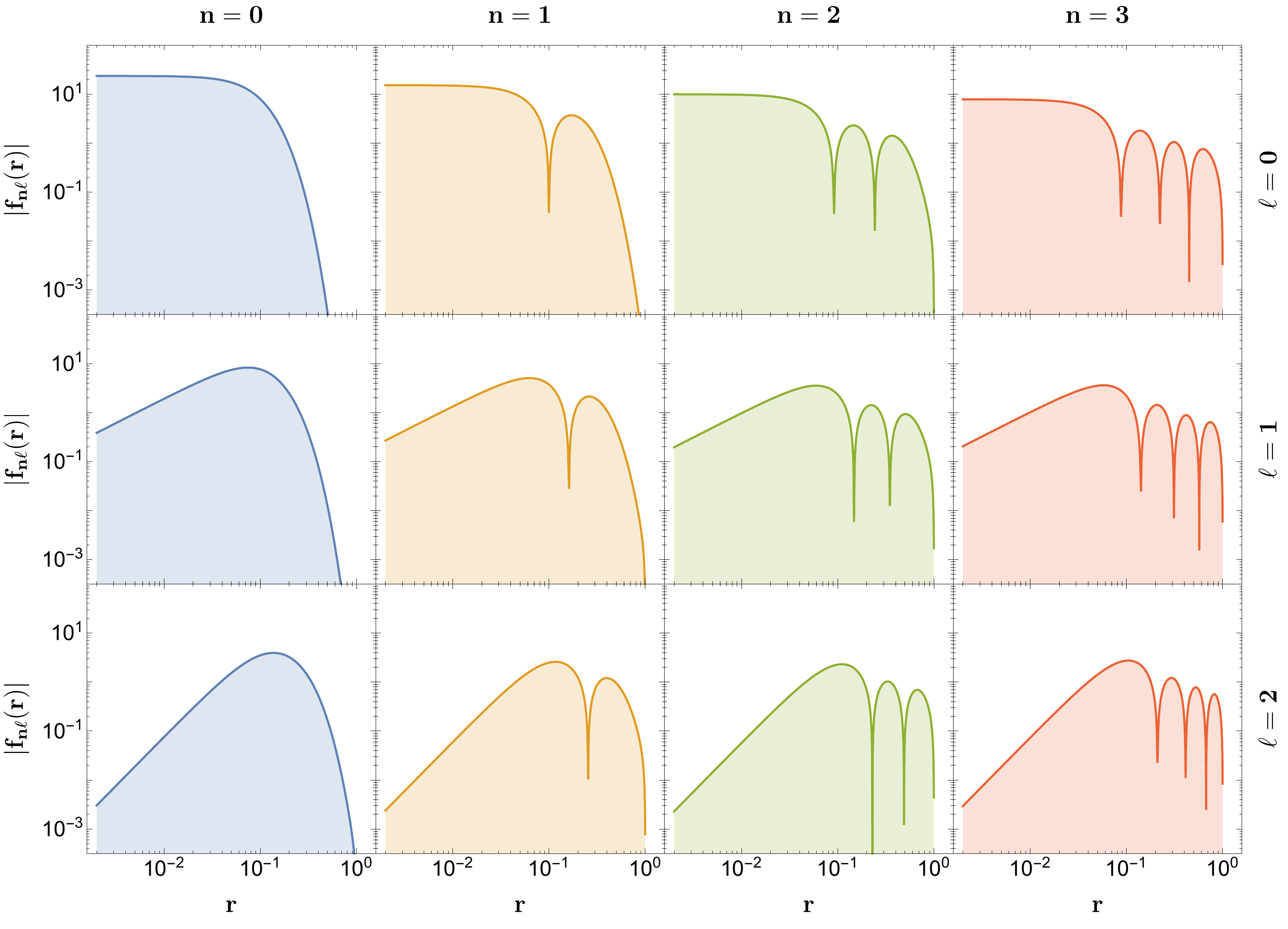}
    \caption{We illustrate the radial profiles $f_{n\ell}$ of the ULDM eigenstates for $n \leq3, \ell \leq 2$. Recall that the $n$-index corresponds to the number of nodes in the state,
      with the energy of the eigenstate increasing with $n$.
      The $\ell$-index corresponds to the angular variation of the wavefunction (given the appropriate $Y_{lm}$); recall that  $f(r) \sim r^{l}$ as $r \rightarrow 0$.
      The $n = 0$ states are colored blue, the $n=1$ states are yellow, the $n=2$ states are green, and the $n=3$ states are red.
      We keep to this convention whenever possible throughout the paper for continuity and clarity. All data is shown in internal code units.
    }
    \label{fig:eigenstates}
\end{figure*}

We now have a formulation of the Schr\"odinger equation that can be solved for a given spherical static potential $\langle \Phi \rangle$. We  discretize our variables into vectors of length $N$  and our operators into  $N$-by-$N$ matrices over a distance $r < r_{\rm max}$ with a grid spacing $\Delta r = r_{\rm max}/N$.
The  differential equation then becomes the matrix eigenvalue problem
\begin{widetext}
\begin{align}
  \left(
  \begin{bmatrix}
    \chi(r_{1}) & \cdots & \cdots & \cdots & 0 \\
    0 & \chi(r_{2})  & \cdots & \cdots  & 0 \\    \vdots & \ddots & \ddots & \ddots & \vdots \\
    0 & \ddots & 0 & \chi(r_{n-1}) & 0 \\
    0 & \cdots & 0 & 0 & \chi(r_{n}) \\
  \end{bmatrix}
  -
   \frac{1}{\Delta r^{2}}
  \begin{bmatrix}
    -2 & 1 & 0 & \cdots & 0 \\
    1 & -2 & 1 & \cdots  & 0 \\
    \vdots & \ddots & \ddots & \ddots & \vdots \\
    0 & \ddots & 1 & -2 & 1 \\
    0 & \cdots & 0 & 1 & -2 \\
  \end{bmatrix}
 \right)
  \begin{bmatrix}
    u_{1} \\
    u_{2} \\
    \vdots \\
    u_{n-1} \\
    u_{n}
  \end{bmatrix}
  =
  2 E_{nl}
  \begin{bmatrix}
    u_{1} \\
    u_{2} \\
    \vdots \\
    u_{n-1} \\
    u_{n}
  \end{bmatrix}
\end{align}
\end{widetext}
where $\chi(r) \equiv 2 \langle \Phi \rangle(r) - l(l+1)/r^{2}$ is the gravitational potential and centrifugal barrier.
This can be solved numerically, with $f_{n\ell} = u_{n\ell}/r$ being the radial component of a given eigenstate and
$E_{n\ell}$ its eigenenergy. 

The boundary conditions must be specified to ensure we have a unique solution. The definition of $u_{n\ell}$ and the requirement that the wavefunction is finite at $r=0$ implies that $u_{n\ell}=0$ at $r=0$. We also
assume the $u_{nl}=f_{nl}=0$ at $\rmax$. Physically, this corresponds to embedding the system in a spherically symmetric infinite well. We clarify the implications of this choice below. Both boundary conditions are built into the matrix equation above.
This outer boundary condition is not the natural choice in a pseudo-spectral code with periodic boundary conditions on a cubic spatial lattice (such as {\sc chplUltra}), but it is easily implemented by setting the wavefunction to zero outside of $\rmax$.

We solve the  matrix equation for a static potential $\langle \Phi \rangle$ corresponding to an unperturbed soliton of mass $M = 50$ in code units.\footnote{We use this as our fiducial ground state in what follows, though our qualitative  results are insensitive to this choice. The FWHM of the soliton is $0.05$ in code units.} The radial $f_{n\ell}$ states that follow from this choice are illustrated in Fig.~\ref{fig:eigenstates}. The $n$-index matches the number of nodes: $n = 0$ states have no nodes, $n=1$ states have one node, and so on. The $\ell$-index is recognizable in the behavior of the function as $r \rightarrow 0$: each state asymptotes to a slope of $r^{\ell}$, such that the $\ell=0$ state has a central core and higher $\ell$-states fall off more quickly.

\subsection{Parameter Dependence of Eigenstates}

\begin{table*}[ht!]
\begin{tabular}{|c|c|rrrrrrrrrrr|}
\hline
$\rmax$ & \multicolumn{1}{c|}{$\langle \Phi (r_{\rm{max}}) \rangle$} & $n = 0$  & $n = 1$  & $n = 2$  & $n = 3$          & $n = 4$  & $n = 5$           & $n = 6$  & $n = 7$  & $n = 8$           & $n = 9$  & $n = 10$ \\ \hline
1.0   & -50                                                 & -406.9 & -175.5 & -93.90 & \textbf{-56.17} & -21.70 & 24.83           & 83.39   & 153.2  & 233.7            & 324.7   & 425.9  \\
2.0   & -25                                                 & -406.9 & -175.5 & -93.91 & -58.05         & -39.34 & \textbf{-27.83} & -16.63 & -2.185 & 15.56           & 36.36  & 60.04  \\
4.0   & -12.5                                               & -406.9 & -175.5 & -93.91 & -58.05         & -39.34 & -28.40          & -21.45 & -16.77 & \textbf{-13.21} & -9.43 & -4.75 \\
8.0   & -6.25                                               & -406.9 & -175.5 & -93.91 & -58.05         & -39.34 & -28.40          & -21.45 & -16.78 & -13.47           & -11.06 & -9.241 \\ \hline
\end{tabular}
\caption{Calculated eigenenergy values in code units for different values of $\rmax$ and $\ell = 0$. The cells where eigenenergies begin exhibiting $\mathcal{O}(1)$ differences from the higher $\rmax$ values are \textbf{bolded}. Note that these also correspond to the appropriate values of the potential at $\rmax$, $\langle \Phi (r_{\rmax}) \rangle$. Thus, comparing the potential at $\rmax\,$ and the derived eigenenergies is a relatively easy way to determine the eigenstates affected by the boundary condition at $\rmax$.
}
\label{tab:eigenenergies}
\end{table*}

When discretizing the Schr\"odinger equation (and subsequently our eigenstates) we made two independent choices: the grid spacing $\Delta r$, and the outer boundary, $\rmax$. Provided $\Delta r$ is small enough to adequately resolve the full width at half maximum (FWHM) of the central soliton, $r_c$, its value does not affect the results of the calculation. We use $\Delta r \approx r_c /25$ throughout.

On the other hand, the value of $\rmax$ qualitatively impacts the eigenstates. Requiring that  the wavefunction vanishes beyond this radius is physically equivalent to putting the entire system into an infinite spherical well of radius $\rmax$.
So long as the radial extent of the eigenfunction is much smaller than $\rmax$ the boundary does not affect our results, but the modes are affected when the scales overlap.
To gain some intuition, let us consider a state with $n$ nodes would fit comfortably into a sphere of some radius $\rmax$. Higher-order states with more than $n$ nodes, then, can only obey the boundary conditions of the same sphere if its nodes are pushed together further than would be the case without the barrier at $\rmax$. The more nodes a state has, the more it is distorted by a boundary at $\rmax$.\footnote{In this work, we consider idealized simulations of a single perturbed soliton or isolated halo in a box, beyond which space is empty, so the wavefunction $\psi$ is effectively zero beyond the boundary.}

Table \ref{tab:eigenenergies} shows eigenenergies for spherically symmetric perturbations ($\ell = 0$) for $n\le10$ and $1<\rmax<10$.  For $n < 2$ these are  identical; at $n = 3$, we see $\mathcal{O}(1)$ differences when $\rmax = 1$. With $n = 5$ we need $\rmax  > 2$ and at $n = 8$ we need $\rmax > 4$ for the eigenenergies to be independent of $\rmax$. Physically,  eigenenergies are independent of $\rmax$ when they do not exceed the (unperturbed) gravitational potential at $\rmax$.

In realistic astrophysical systems, however, 
$\rmax \rightarrow \infty$. We defer a detailed treatment  to future work, but note that for large $\rmax$, the eigenenergies scale as $E_n \sim -1/n^2$, as expected from a hydrogen-like system, until the effect of the spherical well
becomes apparent. This implies a large number of states with relatively small energy splittings near $E \sim 0$.  It is thus possible to excite many of these states as $r_{\rm{max}}\rightarrow \infty$ to similar levels, which could have implications for the relaxation of perturbed solitons to the ground state.

\section{Perturbed Solitons} \label{sec:perturbed-solitons}

We  expand the wavefunction as $\psi = \Sigma_{n \, \ell \, m} f_{n \ell }(r)Y^m_\ell(\theta,\phi)$ and  now explore the time evolution of these states, focusing on perturbations to the gravitational potential that arise as the system evolves. In what follows we fix $m=0$, preserving azimuthal symmetry (although our methods apply to the general case), and write the eigenvectors as $|n \, \ell \rangle$. We focus on perturbing the soliton ground state, $\psi_{\rm{sol}} = f_0(r)Y^0_0(\theta,\phi)$ (or $|n \, \ell \rangle = |0 \, 0 \rangle)$
with  excited states. We construct the eigenstate basis using a gravitational potential with a mass $M=50$ and normalize the eigenstates to unit mass.  

\begin{figure*}[tb]
    \centering
    \includegraphics[width = \textwidth]{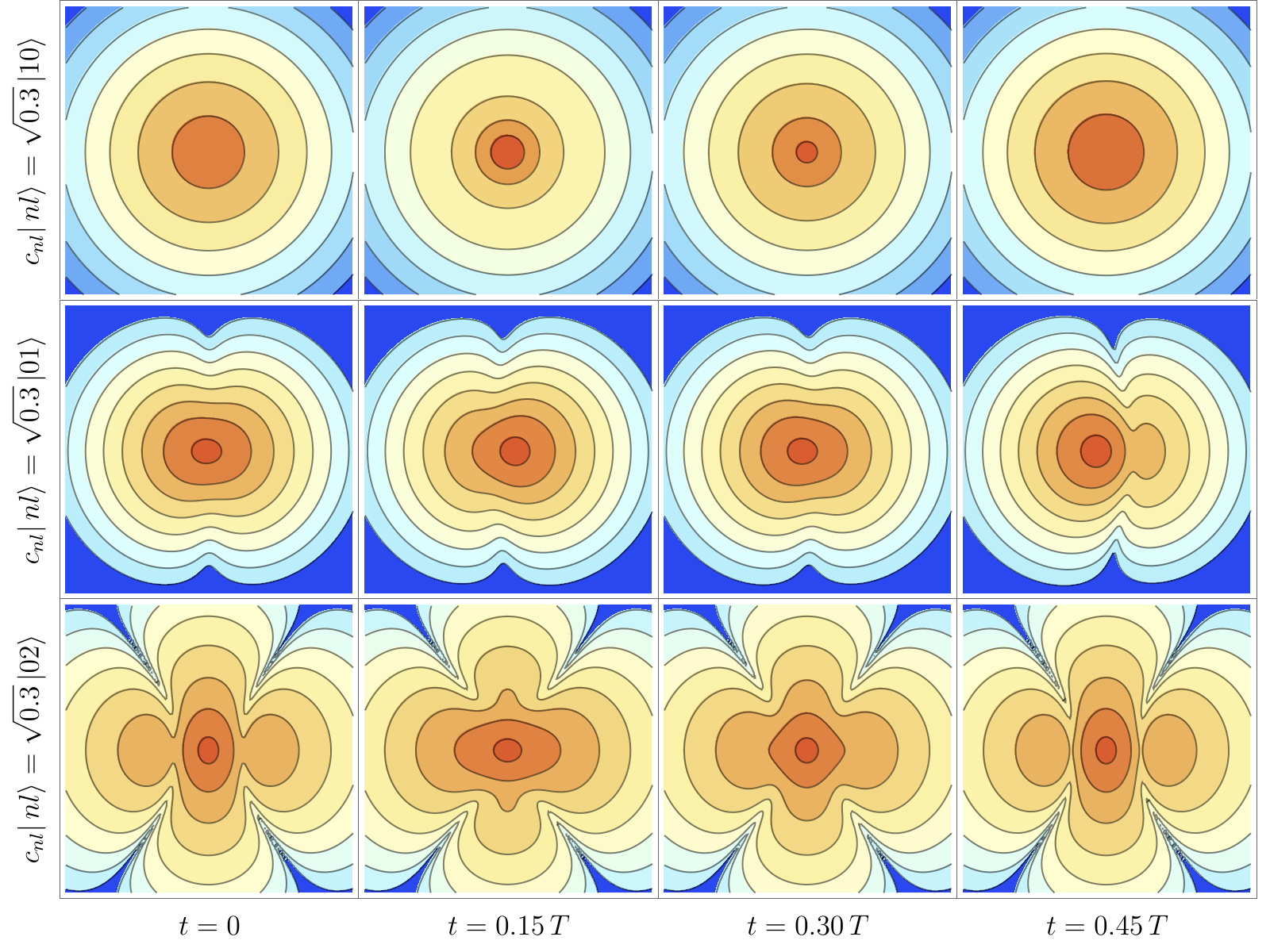}
    \caption{An illustration of how the mass density in the plane is perturbed when combining the soliton ground state with excited states. The perturbations are as follows: row 1 has $30\%$ of its mass in the first $\ell = 0$ excited state, row 2 in the first $\ell = 1$ excited state, and row 3 in the first $\ell = 2$ excited state. Each column represents a time that is defined with respect to the state's period, $T = 2 \pi / \Delta E$. The contours are spaced logarithmically, from $10^{-4}$ to $10^{3}$ in code density units and are kept constant along each row.
    }
    \label{fig:planes}
\end{figure*}

\subsection{Qualitative behavior}

We begin with snapshots of three different systems in which a soliton is perturbed by  $\ket{1 \, 0}$, $\ket{0\, 1}$, and $\ket{0 \, 2}$, shown in Fig.~\ref{fig:planes}. 
In order to illustrate the qualitative behavior of the system, we apply substantial perturbations which induce visible oscillations. In each case the ground state contributes 70\% of the mass density, and the excited $\ell = 0, 1, 2$ states make up the remaining 30\%. Each system is shown at times $t = 0\,T,\, 0.15\,T,\, 0.30\,T,$ and $t = 0.45\,T$ where $T = 2\pi/\Delta E$ is the period of oscillation set by the difference in eigenenergies of the ground state and each perturber.   

The top row of Fig.~\ref{fig:planes} shows the consequence of adding an $\ell = 0$ excited state. This causes the soliton to contract and collapse, revealing the so-called ``breathing mode" that has been noted in ULDM simulations \cite{2019PhRvL.123e1103M}. The $\ell = 1$ mode (middle row) results in the peak of the soliton moving back and forth, in line with Refs.~\cite{2020PhRvL.124t1301S, 2021ApJ...916...27D}, which found that a soliton in a ULDM halo performs a random walk. Finally,  an $\ell = 2$ term (bottom row) results in a quadrupole oscillation, where in the density is elongated first in one direction and then  in the perpendicular direction. These examples illustrate how the phenomenology of ULDM systems overlaps with the eigenstate description, in agreement with Ref.~\cite{2020arXiv201111416L}.

\subsection{Solitons With Spherically Symmetric Perturbations}\label{subsec:ell-eq-0}

We start by examining spherically symmetric systems ($\ell=0$) whose
initial wavefunction is given by
\begin{align}
\psi(t=0) = \sqrt{M} \left( \ket{0} + \epsilon \ket{n} \right)\,,
\end{align}
where we have suppressed the $\ell,m$ indices on the kets for brevity. The unperturbed mass of the system is $M$ while the perturbation increases the mass by
$\epsilon^2 M$ since we are perturbing the wave function and the density scales as $|\psi|^2$ .

\begin{figure*}[tb]
  \centering
  \includegraphics[width=0.9\textwidth]{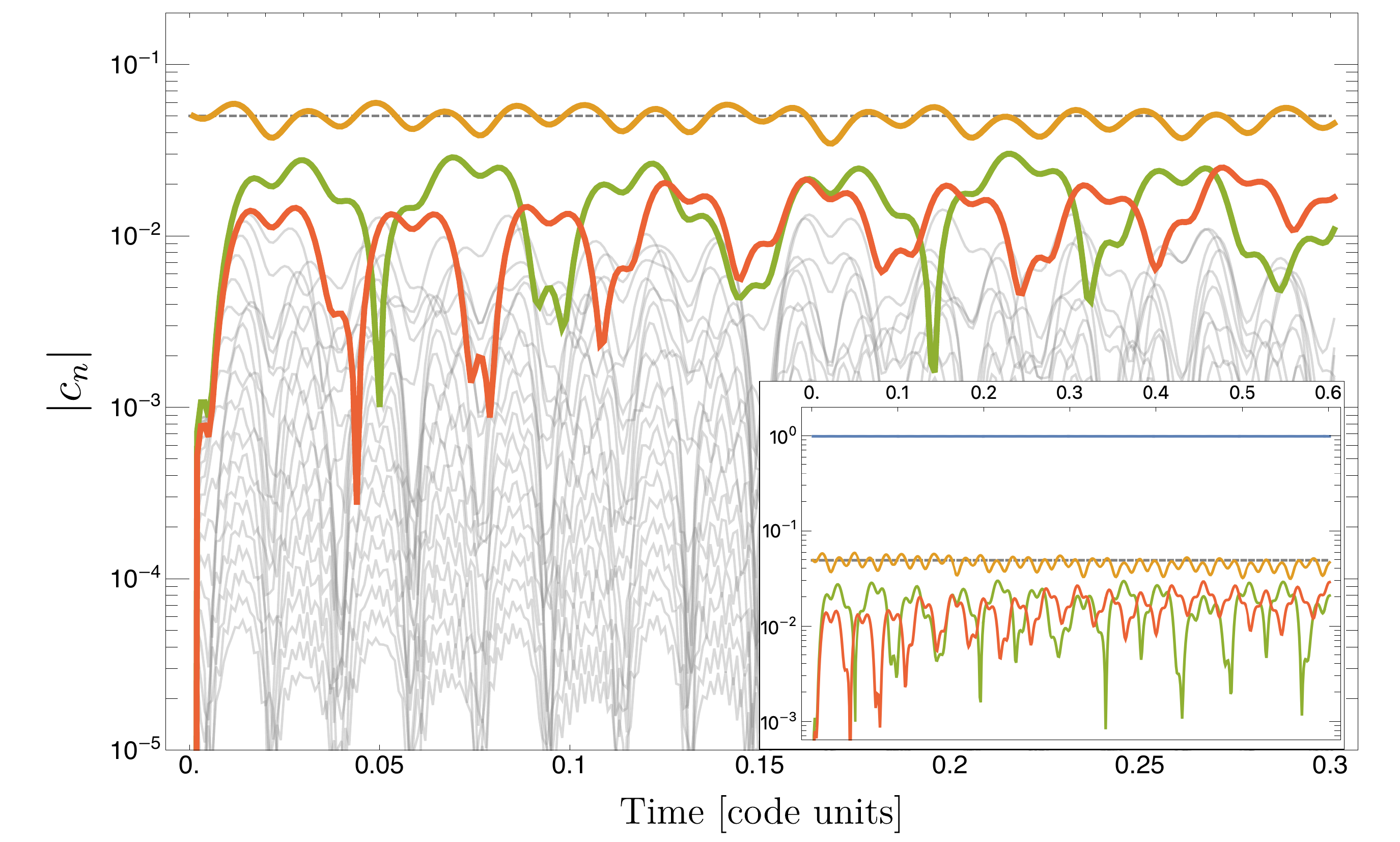}
  \caption{\label{fig:l0n1-time} The evolution of a $M=50$ soliton wavefunction, perturbed by the first $\ell=0$ excited state $\ket{1}$ with amplitude $\epsilon = 5\%$ and expanded into the
  eigenstate basis. The figure shows the magnitudes of these expansion
  coefficients (normalized by $\sqrt{M}$ for the excited states) as a function of time. The prominently
  displayed curves are the amplitudes of the $\ket{1}, \ket{2}$, and $\ket{3}$
  states (from top to bottom), while the other curves show the next 21
  eigenstates. The horizontal dashed line shows the initial amplitude
  $c_1(t=0)=0.05$. The inset shows the same system evolved to a later time, plotted with a lower
  time resolution. Also shown is the evolution of the amplitude of the ground
  state which remains at its initial value of 1 to better than 0.1\%.
}
\end{figure*}

Fig.~\ref{fig:l0n1-time} shows the evolution of the first excited
state ($n=1$) with $\epsilon = 0.05$, drawn from a solution of the full Schr\"{o}dinger-Poisson system. We decompose the full wavefunction into the eigenstate basis $\psi(t) = \sum c_n(t) \ket{n}$ and plot the magnitudes of the $c_n$ with time.  For small perturbations, the amplitude $c_0$ of the ground state will remain constant, and this is true in practice to better than 0.1\% for this scenario. Mode-coupling in the full nonlinear system excites the $\ket{2}$  and $\ket{3}$ modes to significant amplitudes, relative to the original perturbation, as it evolves.   

\begin{figure}[ht]
  \centering
  \includegraphics[width=0.9\columnwidth]{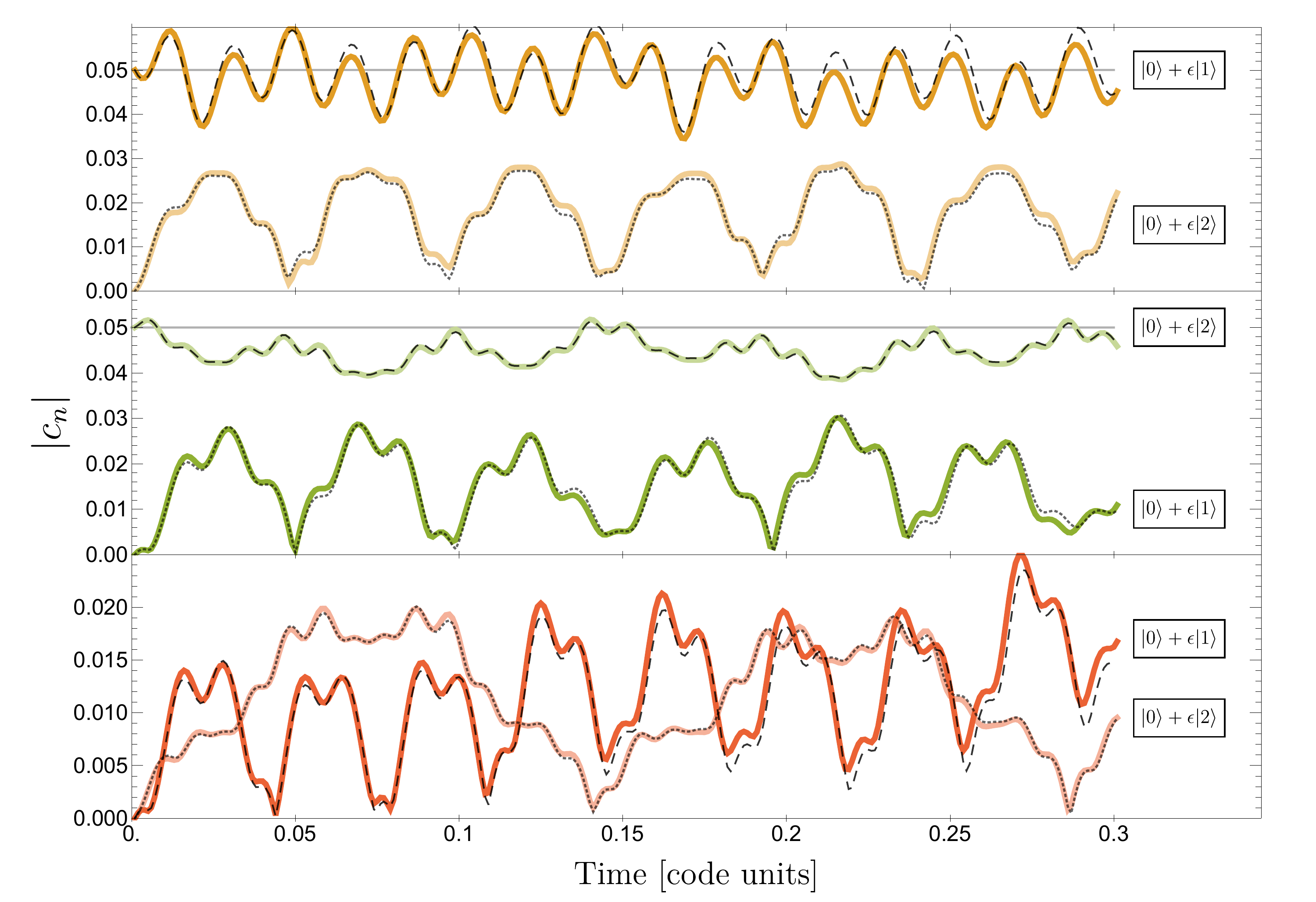}
  \caption{\label{fig:l0n1-time-pt} The time evolution of the amplitudes (from top to bottom)
    of the $\ket{n=1}$, $\ket{n=2}$ and $\ket{n=3}$ eigenstates, compared to a
    perturbative calculation. The brighter color lines show the evolution of states with an initial perturbation proportional to $\ket{n=1}$, while the lighter  lines show the $\ket{n=2}$ case. Perturbative predictions are dashed  and dotted for $\ket{n=1}$ and $\ket{n=2}$ respectively.  
    In the absence of nonlinear couplings due to gravity, the amplitudes would remain at their initial values of $0.05$ and $0$.}
\end{figure}

The  eigenstate expansion  does not account for the gravitational couplings
between modes. To do so, we extend our  expansion to the
interaction picture, 
\begin{align}
  \psi(t) = \sum_{n=1}^{N} c_n(t) \exp \Bigl(-i E_n t\Bigr) \ket{n} \, .
\end{align}
where our expansion coefficients $c_n$ (which are in
general complex) are now time
dependent. The evolving eigenstates will perturb the potential $\Phi \rightarrow \Phi_0
+ \Delta \Phi(t)$,  where $\Phi_0$ is the gravitational potential of the fiducial, ground state profile. The Schr\"{o}dinger equation then reduces to a set of coupled
differential equations for the $c_n$,
\begin{align}
\frac{d c_n}{dt} = -i \sum_{k=0}^N \braket{n|\Delta \Phi |k} c_k(t) e^{-i (E_k-E_n) t} \,\,.
\label{eq:pteqns}
\end{align}
This equation is nominally exact, but also gives a framework with which to approximate
the evolution of this system. To determine $\Delta \Phi$ we first compute
the perturbations to the density profile,
\begin{align}
  \Delta \rho &=  |\psi|^2 - |\psi_o|^2 \\
              & \approx \sum_{p=1}^N 2 \re \left[c_0(t) c_p(t)^{*} \ket{0} \ket{p} e^{i (E_p-E_0) t} \right]
\end{align}
where we drop terms below leading order in $|c_n|$ for $n>0$.
\footnote{Since our Hamiltonian is real and symmetric, it is possible to choose
  our eigenstates to be completely real. We therefore do not need to consider the complex
  conjugate of the eigenstates. We also note the non-standard notation
  $\ket{a} \ket{b} \equiv \psi_{a} \psi_{b}$ for the simple product of eigenfunctions.}
If we define $\Delta \Phi_{0p}$ as the gravitational potential that results from a
density profile $2 \ket{0} \ket{p}$, then Eq.~\ref{eq:pteqns} can be written as
\begin{align}\label{eq:dcndt}
\begin{split}
  \frac{d c_n}{dt} = -i \sum_{p=0}^N\sum_{k=0}^N
    & \braket{n| \Delta \Phi_{0p} |k}  \re \left[c_{0}(t) c_p(t)^{*} e^{i (E_p - E_0)t}\right] \\
    & \times c_{k}(t) e^{-i (E_k-E_n) t} \,\,.
\end{split}
\end{align}
This equation must be slightly modified for $p=0$ to avoid double counting and including the unperturbed solution, but we elide this here for simplicity. We tested the evolving the perturbation equations holding $c_0$ fixed (i.e. ignoring the $p=0$ term) and we find that this makes no difference to our results.

Fig.~\ref{fig:l0n1-time-pt} shows the evolution following  initial perturbations of $\epsilon \ket{1}$ and   $\epsilon \ket{2}$, with $\epsilon=0.05$. In the absence of mode coupling  $\ket{1}$, $\ket{2}$ and $\ket{3}$ would stay at their initial values. We find the perturbative treatment gives a close match to the weights extracted from solutions to the full equations of motion. The discrepancy between the approximation and the full solution grows (albeit slowly) with time.

\begin{figure}[t]
  \centering
  \includegraphics[width=0.9\columnwidth]{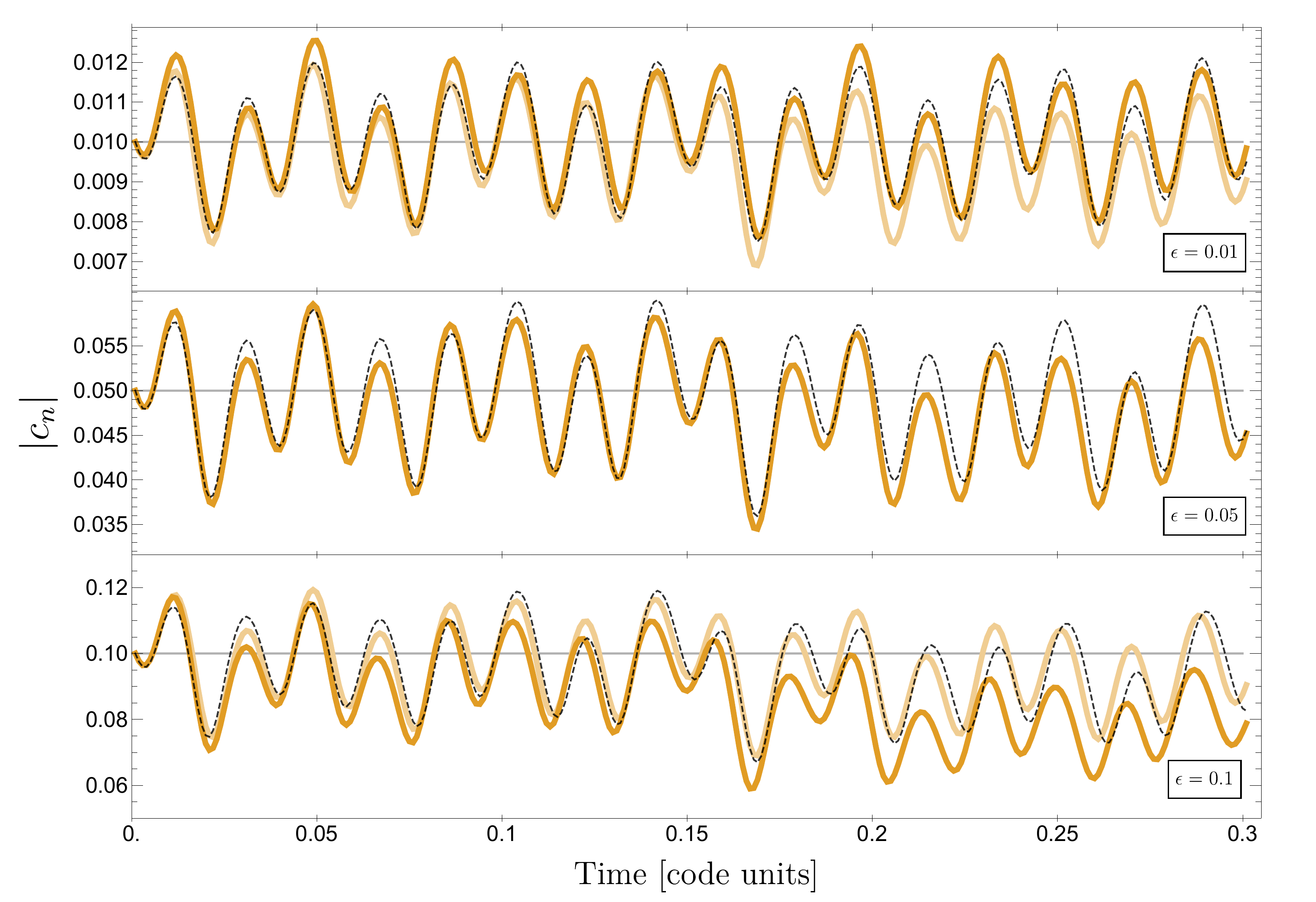}
  \caption{\label{fig:l0n1-ampl-pt} The time evolution of the amplitudes
    of the $\ket{n=1}$ eigenstates compared to a perturbative calculation for different initial amplitudes. The full yellow lines show the evolution of the states when the soliton is perturbed by $\ket{n=1}$ with an initial amplitude of $\epsilon = 0.01$, $\epsilon = 0.05$, $\epsilon = 0.1$, respectively. The lighter yellow lines show a scaled version of the $\epsilon = 0.05$ simulation. The corresponding perturbative calculations are shown in dashed lines.  Note that the $\epsilon = 0.01$ figure has a small range in amplitude, so any divergence between the simulation and perturbation theory is visually amplified and that the perturbative calculation does not match the full system at late times for $\epsilon \geq 0.1$.
  }
\end{figure}

We expect the match between the perturbative calculation and the full system to improve as the  initial amplitude is decreased. Fig.~\ref{fig:l0n1-ampl-pt} demonstrates the expected scaling,  between the simulations; a 10\% perturbation diverges relatively quickly from the full solution, but a 1\% perturbation tracks relatively well through multiple oscillations. As we perturb the soliton with higher energy ($n$) states, we observe that the time-dependence of the resulting $c_n$ amplitude decreases. The amplitude of the 5$^{\rm{th}}$ excited state is constant to within 0.3\%, whilst the 15$^{\rm{th}}$ excited state varies by  0.02\%. It appears that the more rapid fluctuations  in both space and time (higher eigenstates oscillate more rapidly as a function of radius and time) average out variations in the potential, reducing the coupling matrix elements and keeping $c_{n}$ constant in time. This suggests that even when density profiles are composed of many eigenstates,  the lowest order modes dominate the resulting gravitational
couplings and will drive deviations from the simple eigenstate evolution. 

Examining Eq.~\ref{eq:dcndt}, we see that the dominant corrections to a state $\ket{n}$ come from its coupling to the ground state through the potential perturbation, corresponding to the $k=0$ terms. One might expect that these couplings to be further suppressed by the rapidly oscillating exponentials (due to the energy differences). Given this, the largest contribution to the change in $c_n$ comes from the $p=n$ terms. This qualitatively explains why the ground state does not see corrections of order $\epsilon$, but the perturber does, as shown Fig.~\ref{fig:l0n1-ampl-pt}.  

\subsection{Solitons With Aspherical Perturbations}\label{subsec:ell-neq-0}

\begin{figure}[tb]
    \centering
    \includegraphics[width=0.95\columnwidth]{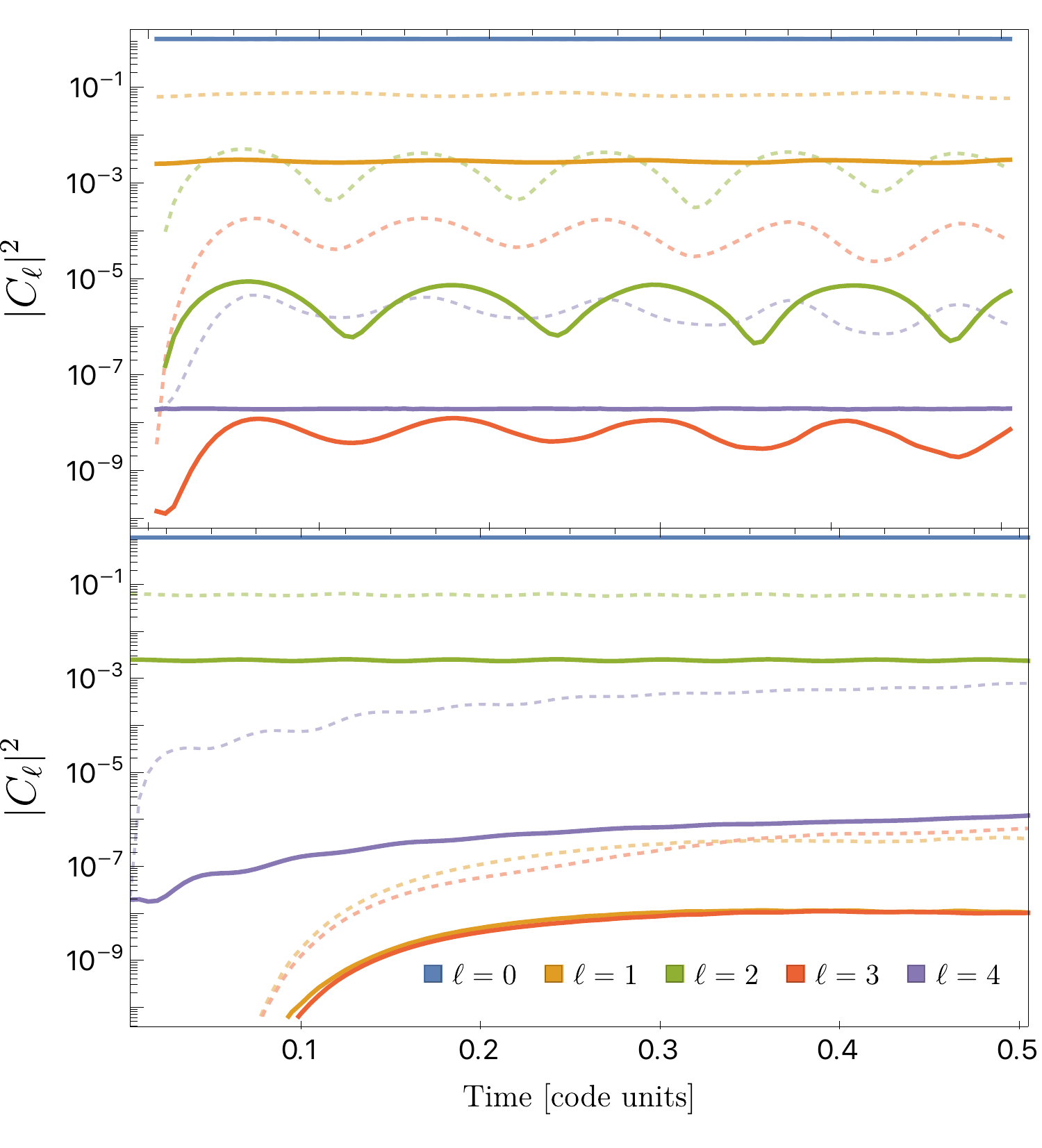}
    \caption{The figure illustrates $|C_{\ell}|^2$ evolution as a function of time in the case of a soliton perturbed by a single non-radially symmetric ($\ell>0$) state. The upper panel illustrates this evolution in the case where the soliton is perturbed by $\ket{0 \, 1}$ with initial amplitudes of $(5\%)^2$  (full lines) and $(25\%)^2$ (dashed lines). The lower panel illustrates the equivalent  $\ket{0 \, 2}$ case. The $\ell$-indices correspond to colors as indicated in the legend at the bottom of the lower panel. Note  the $\ell = 1$ and $\ell = 3$ states in the case of an $\ell = 2$ perturbed (lower panel) are at the noise-floor  in the numerical box.
     } 
    \label{fig:soliton-ell-perturb}
\end{figure}

Next, we turn to full 3D simulations consisting of a single soliton with a nonzero $\ell$-perturbation. Similar to the spherically symmetric
systems, we consider the case
\begin{align}
\psi(t=0) = \sqrt{M} \left( \ket{0 \, 0} + \epsilon \ket{n \, \ell} \right)\,,
\end{align}
where we restore the $\ell$ indices to our kets.\footnote{We continue to set $m=0$.} We use $\ket{n \, 1}$ and $\ket{n \, 2}$ as perturbers for the discussion below, but our conclusions hold for states with higher~$\ell$.
We decompose the
resulting wavefunctions into eigenstates at each saved timestep.
We start by plotting total $\ell$ mode coefficients $|C_\ell|^2 \equiv  \sum_{n} |c_{n,\ell}|^{2}  $ in Fig.~\ref{fig:soliton-ell-perturb}. As with the radial perturbation in Fig.~\ref{fig:l0n1-time-pt} above, the soliton amplitude remains the mostly constant dominant component, while each total $\ell$-mode oscillates about a constant amplitude. The figure shows the mixing between the $\ell$ modes and demonstrates that, to leading order, the $\ell$ modes remain independent of each other. We show that this follows directly from the perturbative treatment below.

In the case where $\ell = 1$ is the initial perturbation of $\sim 5\%$ in the wavefunction, its $|C_\ell|^2$ value oscillates around just above $(5\%)^2 = 0.25\%$, while each subsequent total $\ell$-mode is excited to a progressively smaller amplitude. When $\ell = 2$ is the initial perturbation, each subsequent even value of $\ell$ is excited to a smaller and smaller amplitude, while the odd $\ell$-coefficients are only excited at the level of noise in the simulation box.

We compare these findings with the case where we perturb solitons using the same modes, but at a larger amplitude of $\epsilon = 0.25$. The dominant $\ell $ modes behave
almost exactly the same as in the case of a 5\% perturbation, except that they oscillate around higher amplitudes. On the other hand, by inspecting the higher $\ell$ behavior
we see how the larger perturbation amplitude results in a more pronounced coupling to the higher $\ell$ modes, raising these from noise floor.

Our perturbative treatment from the previous section can be extended to the nonspherical case. As before, we sum over states, except that these now run over
both $\ell$ and $n$, instead of just $n$. We then have
\begin{align}\label{eq:dcnldt}
\begin{split}
\frac{d c_{n_1 \ell_1}}{dt} &= -i \sum_{\ell_2, \ell_3=0}^L \sum_{n_2, n_3=1}^N
\braket{ n_1 \, \ell_1 | \Delta \Phi_{n_2  \ell_2} |n_3\, \ell_3 } \\
& \times \re \left[ c_{0 0}(t) c_{n_2 \ell_2}(t)^* e^{i (E_2 - E_0)t}\right] \\
& \times c_{n_3 \ell_3}(t)  e^{-i (E_3-E_1) t} \, 
\end{split}
\end{align}
where $E_0$ is shorthand the eigenenergy of the unperturbed soliton, $E_1 = E_{n_1 \ell_1}$, $E_2 = E_{n_2 \ell_2}$, and $E_3 = E_{n_3 \ell_3}$ and $L, N$ are the highest $n$ and $\ell$-states
we track.\footnote{The perturbative results for the figures in this manuscript were produced with $N = 25$, $L = 3$.} As before, we approximate the potential perturbations by considering density fluctuations that arise from the combination of the ground state with an excited state.
While the above appears cumbersome, it is identical in structure to the $\ell=0$ case we considered previously. The only new feature comes from the angular terms in the matrix element,
arising from integrating over the product of three spherical harmonics. Appendix~\ref{sec:grav-pot} presents the details of this calculation.

Even without solving these equations, we can recover the qualitative behavior seen in Fig.~\ref{fig:soliton-ell-perturb}. If we work to the
lowest nontrivial order in the perturbation, we see that $n_{3}$
and $\ell_{3}$ must both be zero, i.e. $\ket{n_{3} \, \ell_{3}}$ is the ground state. Considering the product of the three spherical harmonics
in the matrix element $\braket{ n_1 \, \ell_1 | \Delta \Phi_{n_2  \ell_2} |n_3=0 \, \ell_3=0 }$, we see that $\ell_{1} = \ell_{2}$ for
a nonvanishing matrix element at lowest order. Physically, this means that perturbations mix radial eigenstates, but remain at
the same angular eigenstate, which is exactly the behavior seen in the figure. However, this is only true at lowest order\textemdash with larger perturbations there is mixing across angular modes.

\begin{figure}
    \centering
    \includegraphics[width=0.95\columnwidth]{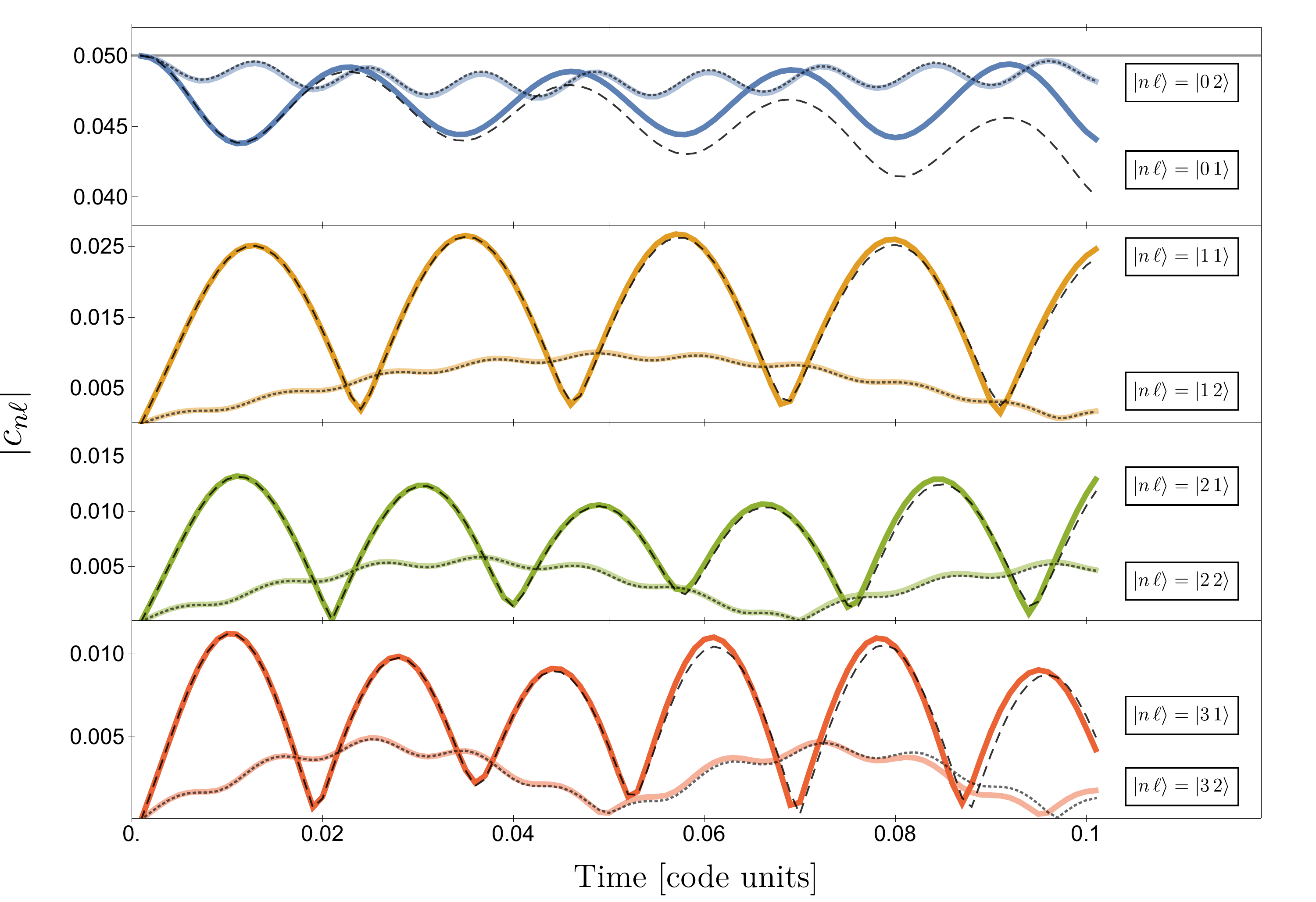}
    \caption{
    The time evolution of the amplitudes (from top to bottom)
    of the $\ket{0 \, \ell}$, $\ket{1 \, \ell}$, $\ket{2 \, \ell}$ and $\ket{3 \, \ell}$ eigenstates for $\ell = 1, \, 2$ compared to a
    perturbative calculation. The brighter color lines show the evolution of the states when the soliton is perturbed by $\ket{0 \, 1}$, while the lighter color lines show a perturbation by $\ket{0 \, 2}$. The perturbative calculations are shown in dashed (for a $\ket{0 \, 1}$ perturber) and dotted (for a $\ket{0 \, 2}$ perturber) lines. 
    }
    \label{fig:l1-l2-PT}
\end{figure}

\begin{figure}
    \centering
    \includegraphics[width=0.95\columnwidth]{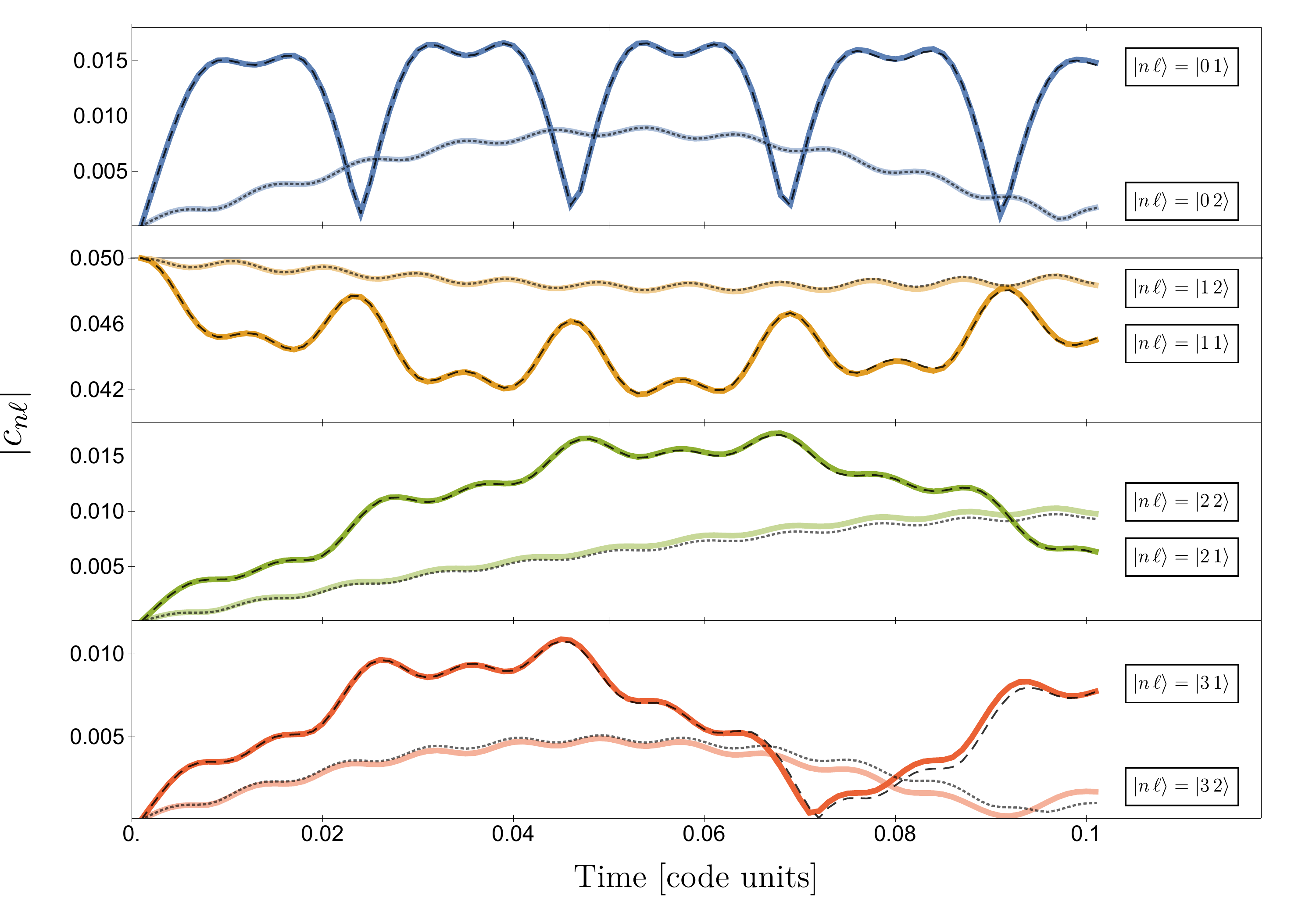}
    \caption{
    The same as Fig.~\ref{fig:l1-l2-PT}, but with the soliton perturbed by $\ket{1 \, 1}$ (darker/dashed) and $\ket{1 \, 2}$ (lighter/dotted). 
    }
    \label{fig:l1n1-l2n1-PT}
\end{figure}

We now proceed by integrating the differential equations as in the previous subsection. The results for perturbing by $\ket{0 \, 1}$ and $\ket{0 \, 2}$ are shown in Fig.~\ref{fig:l1-l2-PT}, and for
perturbing by $\ket{1 \, 1}$ and $\ket{1 \, 2}$ are shown in Fig.~\ref{fig:l1n1-l2n1-PT}. In each of the cases solving Eq.~\ref{eq:dcnldt} accurately matches the evolution of the full system. The perturbative calculation is most accurate for lowest-$n$ states, while at late times higher-$n$ state calculations begin to diverge from simulation data, as is particularly evident in the bottom row of Fig.~\ref{fig:l1n1-l2n1-PT}. We have also verified that the behavior of the system is well captured in the case of a $\ket{0 \, 3}$ perturber, while $\ket{0 \, 4}$ and $\ket{0 \, 5}$ perturbers' values remain constant to better that $0.1\%$, at which level our simulation is subject to noise.

In general, Figs.~\ref{fig:l1-l2-PT} and \ref{fig:l1n1-l2n1-PT} show good agreement between
the simulations and our perturbative calculations. However, one notable divergence is visible in the top row of Fig.~\ref{fig:l1n1-l2n1-PT} for
the soliton perturbed by $\ket{0 \, 1}$.
This highlights a subtlety with our perturbative approach for odd $\ell$ perturbations due to momentum conservation.
The velocity is determined by $d\theta/dx_{i}$ where $\theta$ is the phase of the
wavefunction and $x_{i}$ is a coordinate direction.\footnote{See the Madelung representation of this problem as discussed in eg. Refs.~\cite{2017PhRvD..95d3541H, 2021arXiv210111735H}. }
Consider now a perturbed wavefunction of the form $\ket{00} + c \ket{n \,,{\rm odd\ } \ell}$,
where $c$ is the relative complex amplitude of the perturbation relative to the ground state. If $c$ has a non-zero
imaginary component, the above wavefunction will have a spatially varying phase since the two eigenstates have different shapes.
That, combined with the antisymmetric nature of the odd $\ell$ spherical harmonics, means that the system will have non-zero
overall momentum. For even $\ell$ values, the phase will again be spatially varying, but the net momentum will be zero.

However, the eigenstate expansion does not explicitly conserve the linear momentum of the system.
Structurally, the eigenstate expansion is not translationally invariant and therefore does not have linear
momentum as a conserved quantity.\footnote{By comparison, the eigenstates and perturbation theory are rotationally
  invariant, and so angular momentum is explicitly conserved.}
We can also see this by considering the time evolution of the perturbed wavefunction considered above,
\begin{align}
  \psi(t) = e^{-i E_{0}t} \Big( \ket{00} + c \, e^{-i (E_{n \, \ell}-E_{0})t}\ket{n \,,{\rm odd\ } \ell} \Big) \,.
\end{align}
Even if the imaginary part of $c$ is zero at
$t=0$, the perturbation develops a nonzero relative phase
at a later time, and the system does develop a nonzero momentum (although with a zero
time average value). Interestingly, in our simulations, the relative phase of the $\ket{1 \,0}$ term with the
ground state remains constant at approximately zero, consistent with a vanishing momentum.

While the above suggests an underlying structural problem with any odd $\ell$ mode, Figs.~\ref{fig:l1-l2-PT} and \ref{fig:l1n1-l2n1-PT}
show that significant discrepancies are only evident for the lowest energy $\ell=1$ state. We attribute this to the fact that
this mode generates the largest coherent momentum of the system. Higher energy modes have multiple nodes resulting
in reversals of the velocity direction and higher $\ell$ modes result in a less coherent motion, and therefore a smaller net linear
momentum. Furthermore, while the perturbative theory generically permits coupling across $\ell$ modes, this is not allowed at the lowest order as discussed above. Therefore, even $\ell$ modes do not excite the $\ket{0\,1}$ mode, maintaining good agreement with the perturbative
results.

\section{ULDM Halo} \label{sec:uldm}

We now investigate the eigenstate decomposition and evolution of a ULDM halo. This system can be treated as a solitonic core with an NFW skirt \cite{2020PASA...37....9K}
\begin{align}
\rho(r)=\left\{\begin{array}{ll}
\rho_{\mathrm{sol}}(r), & 0 \leq r \leq r_{\alpha} \\
\rho_{\mathrm{NFW}}(r), & r_{\alpha} \leq r \leq r_{\mathrm{vir}} \, .
\end{array}\right. \label{eq:synthetic-ULDM-halo}   
\end{align}

\begin{figure}
    \centering
    \includegraphics[width=\columnwidth]{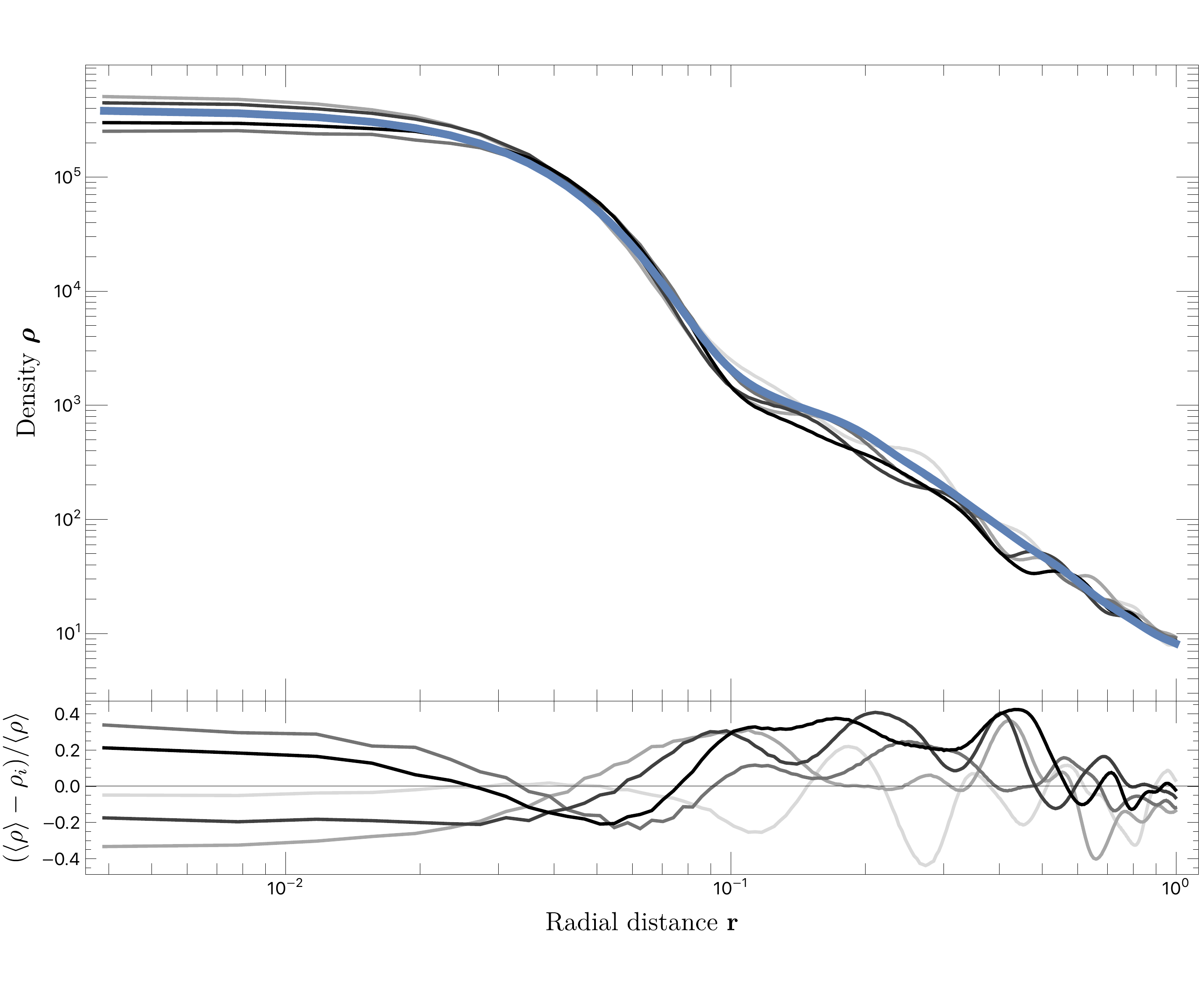}
    \caption{The radially and time-averaged (from $t=0.1$ to $t=1.0$ code units) density profile of our ULDM halo is shown in blue. We use the potential corresponding to this profile to calculate our eigenstates. Snapshots of instantaneous density profiles at $T = 0.2, 0.4, 0.6, 0.8$ and $1.0$ are shown in grayscale (light to dark, respectively). The size of their fluctuations relative to the averaged profile are given in the lower panel. Experimenting with differently time-averaged potentials yielded only small fluctuations in the mass normalization of the resulting eigenstates. All data is shown in code units.
    }
    \label{fig:ULDM-density}
\end{figure}

\begin{figure}
    \centering
    \includegraphics[width=\columnwidth]{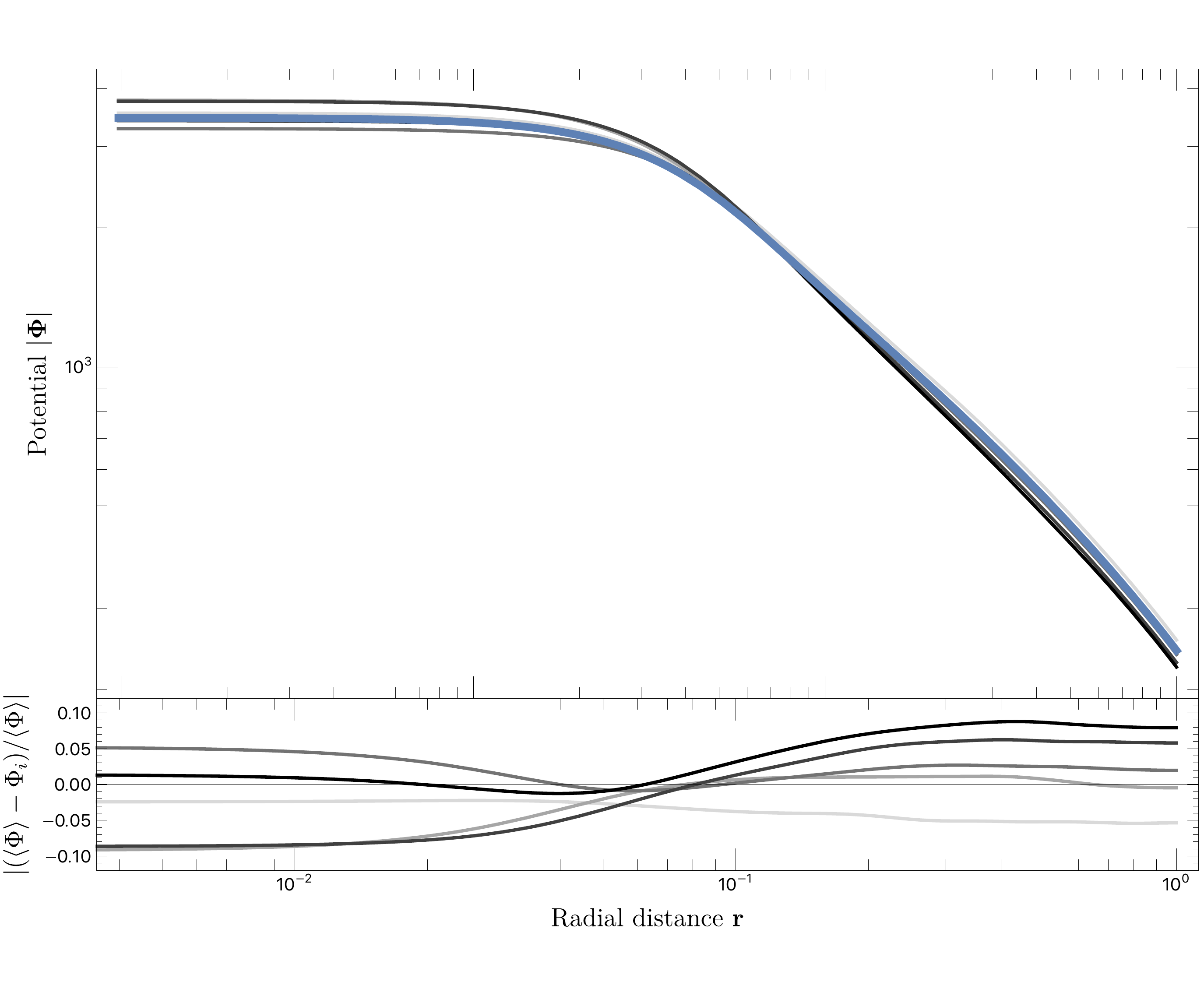}
    \caption{The gravitational potentials corresponding to the density profiles in Fig.~\ref{fig:ULDM-density}. Note that the $\sim40\%$ fluctuations in density correspond to $\sim10\%$ fluctuations in potentials.
    }
    \label{fig:ULDM-potential}
\end{figure}

The border between the skirt and the core falls in the range $3 r_c \leq r_\alpha \leq 4 r_c$, where $r_c$ is the FWHM of the solitonic core and the exact value of $r_\alpha$ is determined by setting the mass of the halo $M_h$ and requiring the profile be continuous. To generate a halo profile that could be described by Eq.~\ref{eq:synthetic-ULDM-halo}, we use {\sc chplUltra} to collide 8 randomly placed equal mass solitons \cite{Schwabe2016}. We then average the resultant late-time profile over $0.9$ code time units. See Fig.~\ref{fig:ULDM-density} for an illustration of our averaged profile compared with instantaneous profiles at different times, and Fig.~\ref{fig:ULDM-potential} for the corresponding potentials.

\begin{figure*}[!ht]
\centering
\includegraphics[width = \textwidth]{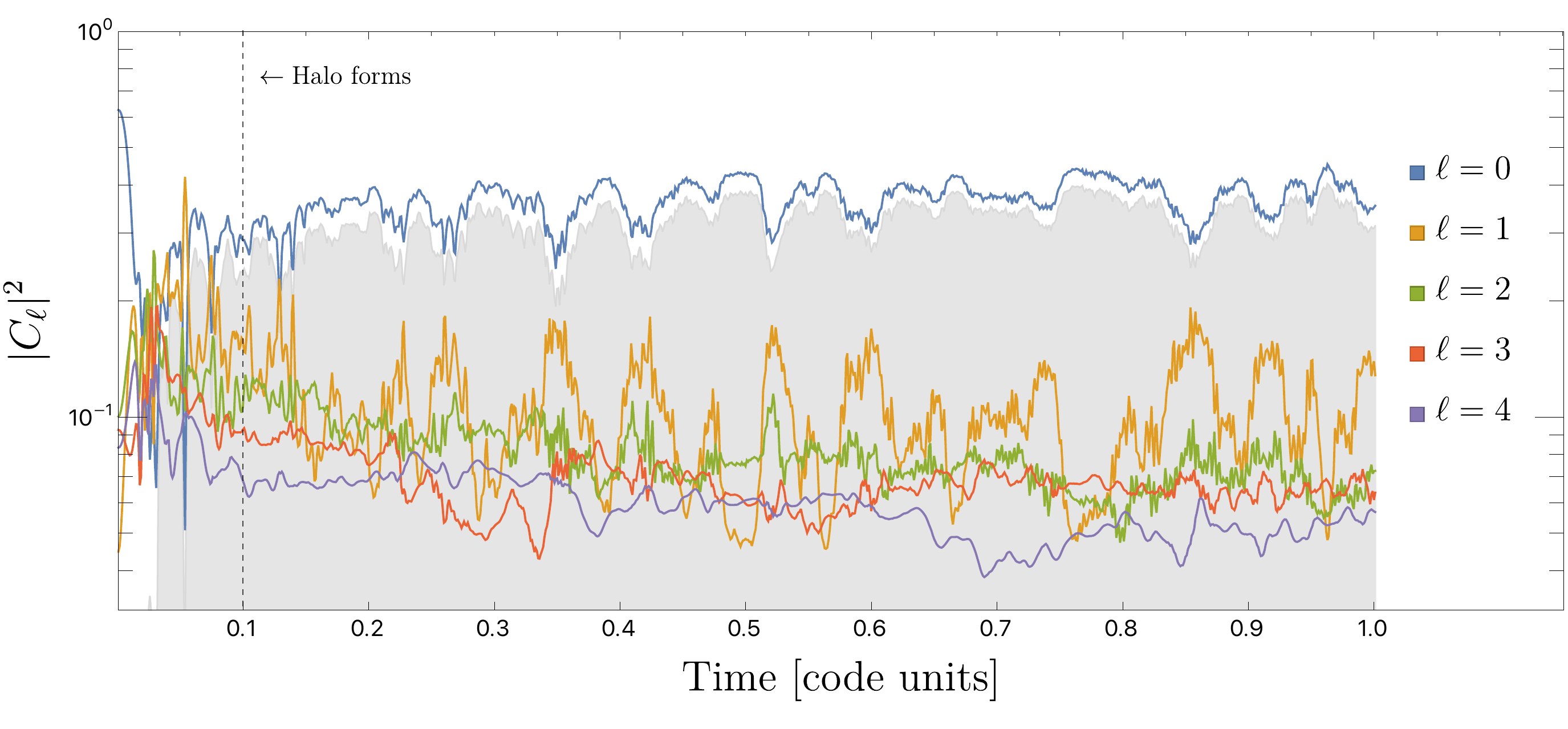}
\caption{The time evolution of $|C_\ell|^2$  in a 3D box with an 8-soliton merger ending in a ULDM profile. The colored lines represent the $\ell$ modes as indicated in the legend and the gray background tracks the evolution of the soliton ($n = 0, \, \ell = 0, \, m = 0$). The vertical line denotes the approximate point of halo formation at time $t = 0.1$.  Note that the $\ell = 0$ line dominates throughout the simulation and is almost entirely composed of the ground state soliton, oscillating around 30\%. The $\ell = 1$ modes make up around 10\% of the halo; the $\ell = 2$ modes make up around 8\%; and higher $\ell$-modes account for around 6\% of the halo wavefunction each.
}
\label{fig:ULDM-Cl}
\end{figure*}

We construct the eigenstates for the potential seeded by the time-averaged ULDM density profile. Next, we analyze the 3D simulation of the 8-soliton collision that led to our profile by decomposing it into its constituent $|C_\ell(t)|^2$ indices.\footnote{In this section we also sum over $m$-modes, as our halo is not axisymmetric and $m \neq 0$ modes contribute significantly.} The results are shown in Fig.~\ref{fig:ULDM-Cl}. At each timestep, the $\ket{00}$ state accounts for the solitonic core at the center of the halo profile, while a superposition of higher modes results in the NFW skirt. We find that the $\ell = 0$ mode dominates, accounting for just over 35\% of the simulated mass, with almost the entirety being in the soliton itself ($\ell = 0$, $n=0$ ). Higher $\ell$-modes account for the halo's NFW skirt, with the $\ell = 1$ making up about 10\% of the wavefunction, albeit with large fluctuations. The $\ell=2$ contributions account for a little more than 8\%, while the $\ell = 3$ and $\ell = 4$ terms account for around 6\% each. The modes presented in Fig.~\ref{fig:ULDM-Cl} account for $\sim 67\%$ of the halo's mass, with the rest being in higher modes.

As in Fig.~\ref{fig:soliton-ell-perturb}, the mean amplitude of each $|C_\ell|^2$ line is roughly constant---albeit with relatively large excursions---suggesting that mass is primarily exchanged between modes with the same $\ell$ number. Mapping to astrophysically reasonable units, the evolution of the system is shown for approximately 23 Gyrs,  the halo mass is $M_h \sim 15 \times 10^8 \, M_\odot$, and its radius is $r_h \sim 20 \rm{kpc}$ (see Table~\ref{tab:units-scaling}). We find no signs of the eigenstate decomposition tending towards a perfectly relaxed state over this time period, even though the density profile of the halo appears to be more stable (as shown in Fig.~\ref{fig:ULDM-density}). It is also possible that this is a result of the artificial construction of this halo, and that the asymmetry in the initial conditions somehow still persists. We plan to explore decompositions for a larger variety of halos in future work.

The relatively large amplitude of non-solitonic modes making up $\sim70\%$ of this halo suggest that our perturbative approximations cannot be applied as simply as in the case of mildly perturbed solitons. In principle, we could attempt to use Eq.~\ref{eq:dcnldt} and significantly increase the $L, N$ cutoff values (i.e., keep track of many more modes) to attempt to find an approximate perturbative match to the full solution. Furthermore, since the differential equations for the time dependent perturbation theory are exact, one could imagine exactly evolving the full system (including a complete calculation of the potential) for a truncated basis. This might provide some advantages over the full Schr\"{o}dinger-Poisson solvers.

\section{Discussion} \label{sec:discussion}

In this paper we solved for the eigenstates and eigenenergies of the Schr\"{o}dinger-Poisson system. We assume that the potential is constant in time, consistent with Ref.~\cite{2020arXiv201111416L}. Once we  obtain the eigenstates of the system, we see phenomena familiar from simulations of ULDM halos. Perturbing the ground state soliton with an $\ell = 0$ component, we recovered the familiar ``breathing mode" exhibited by ULDM solitonic cores; $\ell = 1$ perturbations cause the center of the soliton to move in ways reminiscent of the random walk of the core found in some simulations \cite{2020arXiv201111416L, 2021ApJ...916...27D}; $\ell = 2$ perturbations resulted in a ``cross" oscillation pattern characteristic of the quadrupole moment. We examined the dependence of our eigenstates on the size of our outer boundary condition $\rmax$ and found that higher excited states can be strongly impacted by this choice, but not by our choice of potential.

We tested the accuracy and utility of our perturbative approximation by comparing it with the evolution of the full non-linear Schr\"{o}dinger-Poisson system. We began by comparing the evolution of a radially symmetric system, where the ground state was perturbed by the $|100 \rangle$ state, which we found to be an excellent match when tracking $N \geq 10$ states in our perturbation theory calculation. Additionally, this remains true when the ground state is perturbed with different higher $n$ modes. Finally, we characterized the sensitivity of this approach  to the  perturbation amplitudes, finding that amplitudes in $\psi$ of order 10\% quickly begin to diverge from the full solution but amplitudes of 5\% or less  match.

Extending our perturbation theory calculation to include non-radially symmetric components, we likewise found that  full simulation results match the perturbative prediction. Both of these numerical experiments show that by accounting for the perturbations in the potential, $\Delta \Phi_{j k}$, we were able to achieve a better match between predicted and simulated mode evolution than by simple superposition of modes and their appropriate $e^{-i E_n t}$ evolution used in Refs.~\cite{2020arXiv201111416L, 2021JCAP...03..076D}. The largest divergence between our simulated and perturbative calculations arises because linear momentum is not conserved in our perturbative eigenstate expansion. This effects  only odd $\ell$ modes due to the antisymmetric nature of odd spherical harmonics; furthermore, it is negligible for all except the lowest $\ell = 1$ state, which generates the largest coherent momentum. 

We created a ULDM halo in {\sc chplUltra} by colliding eight randomly placed solitons. We decomposed each snapshot of this simulation into $\ket{n \, \ell}$ eigenstates and tracked the evolution of $|C_\ell|^2$ modes. We found:
\begin{itemize}
    \item the soliton accounts for around $30\%$ of the halo's mass;
    \item higher $\ell=0$ modes account for very little ($\sim5\%$) of the halo mass relative to the soliton;
    \item the $\ell = 1$ modes account for $\sim 10\%$, while $\ell = 2, 3,$ and $4$ account for around 8\% or less each;
    \item the halo does not appear to relax even when evolved over timescales longer than the current age of the Universe. 
\end{itemize}
The relatively large amplitudes of excited modes show that while the perturbative expansion provides insight into the dynamics, fully reproducing its behaviour would require a significant number of terms and accounting for mode-mode interactions.

There are a number of  opportunities created by this work. First, as highlighted by Li {\it et al.\/}
\cite{2020arXiv201111416L}, this eigenstate expansion provides a useful language for describing the evolution
of ULDM systems and a computationally cheap way of synthesizing realistic ULDM halos.
Conversely, this approach has the ability to create benchmark numerical solutions to validate  codes that solve the Schr\"{o}dinger-Poisson system and provides a framework with which to understand the impact that different boundary conditions could have on results. The machinery developed here promises to be useful in analyzing ULDM systems with significant symmetry, such as binary soliton mergers; we will develop this possibility in future work. Moreover, although we  restricted our discussion  to small perturbations of solitons, our approach could form the basis of a simulation tool built around  the time evolution of  a sum of (appropriately designed) eigenstates, as opposed to a spatially discretized wavefunction.  Finally, we speculate that these techniques could provide complementary tools to better understand questions like the mechanisms
by which ULDM systems gravitationally relax and hope to explore these questions in the future.

\begin{acknowledgments}
  We thank Peter Hayman, Lam Hui, Emily Kendall, Xinyu Li, Jens Niemeyer, Victor Robles, and Yourong Frank Wang for useful discussions.
  We thank the Cray/HPE Chapel team, especially Elliot Ronaghan, for collaborating on the development
  of {\sc chplUltra} and for the computational resources used in this paper. This work was performed in part at Aspen Center for Physics, which is supported by National Science Foundation grant PHY-1607611. JZ is supported by the Future Investigations in NASA Earth and Space Science and Technologies (FINESST) grant (award number 80NSSC20K1538). RE acknowledges support from the Marsden Fund of the Royal Society of New Zealand. 
  JZ would further like to dedicate her contribution to this work to the memory of Kosta Pani\'{c}: physicist, teacher, and friend. 
\end{acknowledgments}

\appendix

\section{Simulations with {\sc chplUltra}}\label{sec:code-units}

The simulations use a pseudo-spectral Schr\"odinger-Poisson solver, {\sc chplUltra} \cite{2019SC2}. The algorithm mirrors that of {\sc PyUltraLight}
\cite{2018JCAP...10..027E}, with the added capability to compute the gravitational potential with isolated boundary conditions. We
implement this in {\sc Chapel} \cite{ChapelChapterBalajiBook, ChapelIJHPCA}\footnote{\url{https://chapel-lang.org}}, a next-generation programming language being developed by
Cray/HPE. {\sc Chapel}'s native features allow for productive parallel programming, and (relatively) seamlessly targets systems from traditional
supercomputers to commodity clusters to personal computers. We have successfully scaled {\sc chplUltra} out to 512 nodes, running with grids up to $8192^{3}$,
although most of the results presented in this paper use $512^{3}$ to $1024^{3}$ grids. In addition to {\sc chplUltra}, we also developed a spherically symmetric
code for the $\ell=0$ results. Instead of operator splitting, this directly computes the exponential of a discretized version of the Hamiltonian to implement
the symplectic time stepping. We find good agreement between runs done with both codes.

All of our results are presented in ``code'' units. To convert these to more astrophysically recognizable values, we start by recalling that the Schr\"{o}dinger-Poisson system remains invariant when scaled by a parameter $\lambda$ as follows \cite{PhysRevD.50.3655}:
\begin{equation}
\{t, x, V, \psi, \rho\} \rightarrow\left\{\lambda^{-2} \hat{t}, \lambda^{-1} \hat{x}, \lambda^{2} \hat{V}, \lambda^{2} \hat{\psi}, \lambda^{4} \hat{\rho}\right\}
\end{equation}

From the above, we can calculate how the total mass, energy, and angular momentum scale with $\lambda$:
\begin{equation}
\{M, E, L\} \rightarrow\left\{\lambda \hat{M}, \lambda^{3} \hat{E}, \lambda \hat{L}\right\}
\end{equation}

Furthermore, the Schr\"{o}dinger-Poisson system can also be transformed through scaling the ULDM particle mass $m_a \rightarrow \alpha m_a$ as:
\begin{equation}
\{t, x, V, \psi, \rho\} \rightarrow\left\{\hat{t}, \alpha^{-1/2} \hat{x}, \alpha^{-3/2} \hat{V}, \alpha^{-1} \hat{\psi}, \alpha^{-3/2} \hat{\rho}\right\}
\end{equation}

with the total mass, energy, and angular momentum then scaling as
\begin{equation}
\{M, E, L\} \rightarrow\left\{\alpha^{-3/2} \hat{M}, \alpha^{-5/2} \hat{E}, \alpha^{-2} \hat{L}\right\} \, .
\end{equation}

We adopt a fiducial value of $m_{a} = m_{22} \times 10^{-22}  \, {\rm eV}$, where the scaling of our results with the axion mass is captured by $m_{22}$. Finally, we can introduce appropriate length, time, and mass scales as in Ref.~\cite{2018JCAP...10..027E} as a function of the parameters $\lambda$ and $m_{22}$:
\begin{align}
\mathcal{L} &=\left(\frac{8 \pi \hbar^{2}}{3 m_a^{2} H_{0}^{2} \Omega_{m_{0}}}\right)^{\frac{1}{4}} \approx 38.3 \, \mathrm{kpc} \times \lambda^{-1} m_{22}^{-\frac{1}{2}}, \\
\mathcal{T} &=\left(\frac{8 \pi}{3 H_{0}^{2} \Omega_{m_{0}}}\right)^{\frac{1}{2}} \approx 75.5 \, \mathrm{Gyr} \times \lambda^{-2}, \\
\mathcal{M} &=\left(\frac{8 \pi G^4}{3 H_{0}^{2} \Omega_{m_{0}}}\right)^{-\frac{1}{4}}\left(\frac{\hbar}{m_a}\right)^{\frac{3}{2}} \approx 2.2 \times 10^{6} \, \mathrm{M}_{\odot} \times \lambda m_{22}^{-\frac{3}{2}}.
\end{align}

Each of these scales is equal to one code unit of length, time, and mass, respectively. We present a few choices of $\lambda$ for different astrophysical systems in Table~\ref{tab:units-scaling}.

%
%

\begin{table}[t]
\centering
\begin{tabular}{l|c|ccc|}
\cline{2-5}
                                             & \multirow{2}{*}{$\boldsymbol{\lambda}$} & \textbf{t  {[}Gyr{]}} & \textbf{x  {[}kpc{]}}           & \textbf{M} {[}$\boldsymbol{M_{\odot}}${]} \\
                                             &                                     & $\lambda^{-2}$        & $\lambda^{-1} \, m_{22}^{-1/2}$ & $\lambda \, m_{22}^{-3/2}$            \\ \hline
\multicolumn{1}{|l|}{ Units in Ref.~\cite{2018JCAP...10..027E}}     & $ 1.0 $                             & $75$                  & $ 38.3$                         & $2.2 \times 10^{6}$                   \\
\multicolumn{1}{|l|}{One gigayear time unit}           & $8.7$                               & $1$                   & $4.4 $                          & $1.9 \times 10^7 $                    \\
\multicolumn{1}{|l|}{Hubble time unit}       & $ 2.3$                              & $14$                  & $16.6 $                         & $5.1 \times 10^6$                     \\
\multicolumn{1}{|l|}{Dwarf galaxy halo core} & $ 1.8$                              & $ 23$                 & $21.2 $                         & $4.0  \times 10^{6}$                  \\
\multicolumn{1}{|l|}{Very massive halo core} & $4500$                              & $4 \times 10^{-6}$    & $0.008 $                        & $1.0  \times 10^{10}$                 \\ \hline
\end{tabular}
\caption{
This table lists the physical values corresponding to a single code unit of time, length and mass. We consider different physical situations with their corresponding $\lambda$ values. We also highlight the scaling of these values with $\lambda$ and the mass of the axion on the top, but have only considered $m_{22} = 1$ in the construction of this table. }
\label{tab:units-scaling}
\end{table}


\section{Calculating the gravitational potential}\label{sec:grav-pot}

We require the gravitational potential $\Phi$ from densities of the form $\rho_{lm}(r) Y_{lm}(\theta, \phi)$
\begin{align}
  \nabla^{2} \Phi = 4 \pi \rho_{lm}(r) Y_{lm}(\theta,\phi) \,,
\end{align}
where we assume that the potential vanishes at infinity. Recalling that the spherical harmonics are eigenfunctions of the angular Laplacian, the solution must have the form $\Phi = \Phi^{\rm rad}(r) Y_{lm}$.
Making the change of variables $y=r \Phi^{\rm rad}$, the radial part of Poisson's equation becomes
\begin{align}
  \frac{\partial^{2} y}{\partial r^{2}} - \frac{\ell (\ell+1)}{r^{2}} = 4 \pi r \rho_{lm}(r)
\end{align}
with boundary conditions
\begin{align}
    y&(r=0) = 0\\
  y&(r_{\textrm{max}}) = -\frac{4\pi}{2 \ell + 1} \frac{1}{r_{\rmax}^{\ell}} \int_{0}^{\rmax} dr' r'^{2} \rho_{lm}\, ,
\label{eq:upper}
\end{align}
where the upper boundary condition follows directly from the Laplace expansion of the Green's function for a  $1/r$ potential,
assuming that the density has vanished by $\rmax$. Note that for $\ell=0$, the
upper boundary condition is simply $y=-M$ where $M$ is the total mass, as expected for a spherically symmetric problem. We solve this by rewriting the differential equation
as a linear algebra problem, similar to our treatment of the Schr\"{o}dinger equation. Note that we could have just as easily just used the Green's function, but
we find the linear algebra approach more convenient computationally.

Given the potential, we are able to calculate its expectation value with any two other states as follows:
\begin{align*}
\langle j | \Delta \Phi_{0p} | k \rangle &= \int dr d\Omega (f_j^* Y_j^*) (\Delta \Phi_{0p}^{\rm{rad}} Y_0^* Y_p) (f_k Y_k)\\
&= (4\pi)^{-1/2} \int dr f_j^* \Delta \Phi_{0p}^{\rm{rad}} f_k \int d\Omega  Y_j^* Y_p Y_k \, ,
\end{align*}
where we used the shorthand $j = n_1 \ell_1$ and $k = n_2 \ell_2$ when comparing to Eq.~\ref{eq:dcnldt}. Here, we are using $\Delta \Phi_{0p}^{\rm{rad}}$ to refer to the radially-dependent piece of the potential arising from the product of state $p$ with the ground state, while its spherical behavior is captured by the two spherical harmonics. Thus, we can split the integration into the radial piece (which is the same as the spherically symmetric case in Sec.~\ref{subsec:ell-eq-0}) and a new aspherical piece. Performing the replacement $Y_0 = (4 \pi)^{-1/2}$ our angular piece becomes an integral over three spherical harmonics, equivalent to a Wigner 3j symbol \cite{wigner1993matrices}.

\bibliographystyle{apsrev4-2}
\bibliography{bibliography.bib}

\end{document}